\documentclass[letter,11pt]{article}
\usepackage{pos}
\usepackage{graphicx,rotating,wrapfig}
\usepackage{amsmath,amssymb}
\usepackage{hyperref}
\usepackage{xcolor} 
\definecolor{nicered}{rgb}{0.5,0.,0.}
\definecolor{nicegreen}{rgb}{0.,0.5,0.}
\definecolor{niceblue}{rgb}{0.,0.,0.5}
\definecolor{darkpink}{rgb}{0.8,0.47,0.47}
\hypersetup{colorlinks,citecolor=nicegreen,linkcolor=nicered,urlcolor=niceblue}

\title{The upcoming CTEQ-TEA parton distributions in a nutshell}

\author{
  A. Ablat,$^1$ A. Courtoy,$^2$ S. Dulat,$^{1,6}$ M. Guzzi,$^3$ T.~J.~Hobbs,$^4$ T.-J. Hou,$^5$\\
  J. Huston,$^6$ P.~Nadolsky,$^{6,7}$\footnote{Speaker} I. Sitiwaldi,$^1$ K. Xie,$^6$ C.-P. Yuan$^6$}

\affiliation{
			$^1$Xinjiang University, Urumqi, Xinjiang 830046 China\\
			$^2$IF UNAM, Apartado Postal 20-364, 01000 Ciudad de M\'exico, Mexico \\
			$^3$Kennesaw State University, Kennesaw, GA 30144, USA \\
			$^4$High Energy Physics Division, Argonne National Laboratory, Lemont, IL 60439, USA \\
			$^5$University of South China, Hengyang, Hunan 421001, China\\
			$^6$Michigan State University, East Lansing, MI 48824, USA \\
			$^7$Southern Methodist University, Dallas, TX 75275-0181, USA\\ 
		}

\emailAdd{nadolsky@smu.edu}

\abstract{We review recent studies by the CTEQ-TEA group toward the development of a new generation of precision parton distribution functions in the nucleon for advanced studies at the high-luminosity LHC and in other experiments. Among several ongoing efforts, we 
examine sensitivity to the PDFs and mutual compatibility of new measurements in production of Drell-Yan pairs, top-quark pairs, and single-inclusive jets by the ATLAS, CMS, and LHCb collaborations in the LHC Runs 1 and 2.}

\FullConference{31st International Workshop on Deep Inelastic Scattering (DIS2024)\\
 8–12 April 2024\\
Grenoble, France\\}

\begin{document}
\maketitle
Accurate parametrizations for parton distribution functions (PDFs) in the nucleon are essential for a wide range of measurements pursued at the Large Hadron Collider (LHC) and in other experiments. The CTEQ Tung Et Al. (CTEQ-TEA) group is involved in the global analysis of QCD measurements across a large energy span aimed at determination of PDFs at the precision frontier. As the LHC enters its high-luminosity decade, the knowledge of PDFs becomes even more central for accuracy control in many tests of the Standard Model (SM) and searches for beyond-SM physics. At the DIS'2024 workshop, the CTEQ-TEA group reported on a program of theoretical and methodological developments leading to a new generation of PDFs for advanced studies at the LHC Run-3. The new PDFs will replace the widely used CT18 PDFs \cite{Hou:2019efy} in general-purpose and specialized applications. Given the elevated requirements for accuracy of central PDFs and quantification of PDF uncertainties, the development of the new PDF series takes several years and involves implementation of new advances in theoretical calculations, adoption of new data sets constraining the PDFs, and refinement of the fitting methodology.

During the last year, our group published several articles focusing on specific aspects of our PDF fits. In addition, we recently published a summary \cite{Ablat:2024muy} that highlights the key outcomes from the totality of our recent publications and emphasizes the synergy of all ongoing CTEQ-TEA efforts toward obtaining accurate, comprehensive, and reliable future PDFs. Traditionally, theoretical developments involve implementation of radiative contributions to improve the perturbation theory. With the growing availability of perturbative hard cross sections at N3LO in QCD, there is also increasing interest in producing N3LO PDFs. The CTEQ fitting package already includes components of the N3LO PDF analysis, notably the complete quark flavor infrastructure \cite{WangBowenThesis,Xie:2019thesis} for implementation of 3-loop radiative contributions in DIS with massive charm and bottom quarks in the SACOT-MPS general mass scheme and for evolution of PDFs at N3LO accuracy. However, many more components at N3LO are needed to guarantee the N3LO accuracy of the PDF fits and will not be available for a while. PDF-fitting groups~\cite{Cooper-Sarkar:2024crx}, including CTEQ-TEA, implement the mandatory components as they become available. An additional consideration is that, when perturbative uncertainties are suppressed by including appropriate radiative contributions, other types of uncertainties become prevalent in the full PDF uncertainty. Efforts to control these other uncertainties must complement implementation of N3LO contributions.

The upcoming CTEQ-TEA analysis will therefore release NNLO PDFs as well as some investigations done at partial N3LO -- the highest order of QCD available for PDF fits at the moment. The summary article \cite{Ablat:2024muy} reviews the multi-prong efforts toward this goal: 
investigations of the impact of candidate data sets from lepton pair \cite{Sitiwaldi:2023jjp}, top-quark pair \cite{Ablat:2023tiy}, and inclusive (di)jet production \cite{Ablat:2024Jet} from the LHC at 5, 7, 8, and 13 TeV;
advances in methodology to quantify the mutual agreement of experimental constraints \cite{Jing:2023isu,Kotz:2024dfg,Yan:2024yir} 
and streamline estimations of uncertainties due to PDF parametrizations \cite{Kotz:2023pbu, Kriesten:2023uoi,Kriesten:2024are};
studies of small-$x$ dynamics that affects the gluon and other PDFs at all $x$ via sum rules, as well 
as forward charm production \cite{Xie:2021ycd} and high-energy neutrino scattering
\cite{Xie:2023suk}; exploration of first lattice QCD constraints on strangeness quark-antiquark asymmetry \cite{Hou:2022onq}; 
NNLO PDF fits with contributions from nonperturbative (power-suppressed) charm quarks \cite{Guzzi:2022rca} and photons in the proton \cite{Xie:2021equ} and neutron \cite{Xie:2023qbn}; investigations of implications and future prospects ranging from low-energy parity-violating DIS \cite{Accardi:2023chb} to combined PDF-SMEFT fits \cite{Gao:2022srd}.

\begin{table}[tb]
	\caption{ The numbers of points ($N_{\rm pt}$) and $\chi^2/N_{\rm pt}$ values for the LHC Drell-Yan, $t\bar t$ and inclusive jet data sets fitted simultaneously in the CT18+nDYTTIncJet NNLO fit. The uncertainties are obtained with the final Hessian error sets at 90\% CL. The $\chi^2/N_{\rm pt} $ values in parentheses are obtained with the original CT18 NNLO error sets.}\label{Tab:LHCnewData}
	\centering
	\begin{tabular}{clcccccc}
		\hline
		ID & Experiment & $N_{\rm pt}$ &  CT18+nDYTTIncJet (CT18)& \\
		\hline            
		\multicolumn{5}{c}{Drell-Yan pair production} \\
		\hline
		211  & ATLAS 8 TeV W      & 22 & $2.42^{+2.49}_{-1.51}$   ( $4.25^{+6.39}_{-3.34}  $ )  \\
        212  & CMS 13 TeV Z       & 12 & $2.48^{+4.76}_{-0.88}$   ( $12.03^{+38.04}_{-21.84}$)  \\
        214  & ATLAS 8 TeV Z3D    & 188& $1.12^{+0.46}_{-0.02}$   ( $1.99^{+5.10}_{-1.85}  $ )  \\
		215  & ATLAS 5.02 TeV W,Z & 27 & $0.82^{+0.55}_{-0.16}$   ( $1.15^{+1.22}_{-0.43}  $ )  \\
		217  & LHCb 8 TeV W       & 14 & $1.35^{+0.59}_{-0.61}$   ( $1.35^{+0.72}_{-0.64}   $)  \\
		218  & LHCb 13 TeV Z      & 16 & $1.18^{+1.42}_{-0.60}$   ( $1.49^{+1.74}_{-0.89}   $)  \\
		\hline
		\multicolumn{5}{c}{ $t\bar{t}$ production at 13 TeV} \\
		\hline
		521 & ATLAS all-hadronic $y_{t\bar{t}}$ & 12 & $1.06^{+0.14}_{-0.09}$ ($1.05^{+0.21}_{-0.10}$) \\
		528 & CMS dilepton  $y_{t\bar{t}}$ & 10 & $1.10^{+1.08}_{-0.68}$ ( $1.03^{+1.60}_{-0.74} $ )   \\
		581 & CMS lepton+jet  $m_{t\bar{t}}$ & 15 &  $1.44^{+1.18}_{-0.73}$ ( $1.37^{+1.86}_{-0.82}$ )  \\
        587 & ATLAS lepton+jet  $m_{t\bar{t}}+y_{t\bar{t}}+y^B_{t\bar{t}}+H_T^{t\bar{t}}$ & 34 &  $0.92^{+0.32}_{-0.14}$ ( $0.94^{+0.59}_{-0.16}$ )   \\
		\hline
		\multicolumn{5}{c}{Inclusive jet production} \\
		\hline
		553 & ATLAS 8 TeV IncJet  &171 & $1.76^{+0.20}_{-0.12}$ (  $1.80^{+0.33}_{-0.16}$ )   \\
		554 & ATLAS 13 TeV IncJet  &177 & $1.38^{+0.13}_{-0.10}$ (  $1.39^{+0.20}_{-0.11}$ )   \\
		555 & CMS   13 TeV IncJet  &78  & $1.10^{+0.24}_{-0.17}$ (  $1.11^{+0.30}_{-0.16}$ )  \\
		\hline
      & Total &  4457 &  5402\\
   \hline
	\end{tabular}
\end{table}

In this contribution, we highlight one aspect of the ongoing analysis: investigation of the likely impact of new precision data sets from the LHC Runs 1 and 2 on the upcoming PDFs. In Refs.~\cite{Sitiwaldi:2023jjp,Ablat:2023tiy,Ablat:2024Jet}, we examined constraints on NNLO PDFs using measurements from 20 publications on Drell-Yan pair, top-antitop, inclusive jet, and jet pair production published by the ATLAS, CMS, and LHCb Collaborations. Most of these publications provide several differential distributions that may be suitable for the PDF fits. The above publications find that the considered distributions may have non-identical preferences for PDFs because of the impact of collateral systematic factors that are not the same across various measurements. One important case is the gluon PDF at momentum fractions of $x\approx 0.05$ and QCD scales close to 125 GeV that controls Higgs boson production rates via gluon-gluon fusion. 
Sensitivity analyses like the one in \cite{Jing:2023isu} indicate that the main relevant constraints on the gluon PDF arise from several types of processes: scaling violations in DIS at HERA, BCDMS, and NMC; hadron jet production; and, as experimental precision grows, even from Drell-Yan and $t\bar t$ pair production data sets. Ref.~\cite{Ablat:2024muy} pointed out that the relevant high-precision LHC data sets generally exert opposing pulls: in the considered scenarios, Drell-Yan pair and $t\bar t$ pair production on the whole preferred a smaller gluon PDF at $x\gtrsim 0.05$, while inclusive jet production preferred a larger gluon at $x\gtrsim 0.1$. We also compared the PDF constraints from Run-1 and 2 jet production measurements presented either as distributions of single-inclusive jets or distributions of jet pairs. We found that the PDF constraints from single-inclusive jet distributions were less affected by the QCD scale dependence than the counterpart dijet ones. 

\begin{figure}[tb]
	\includegraphics[width=0.99\textwidth]{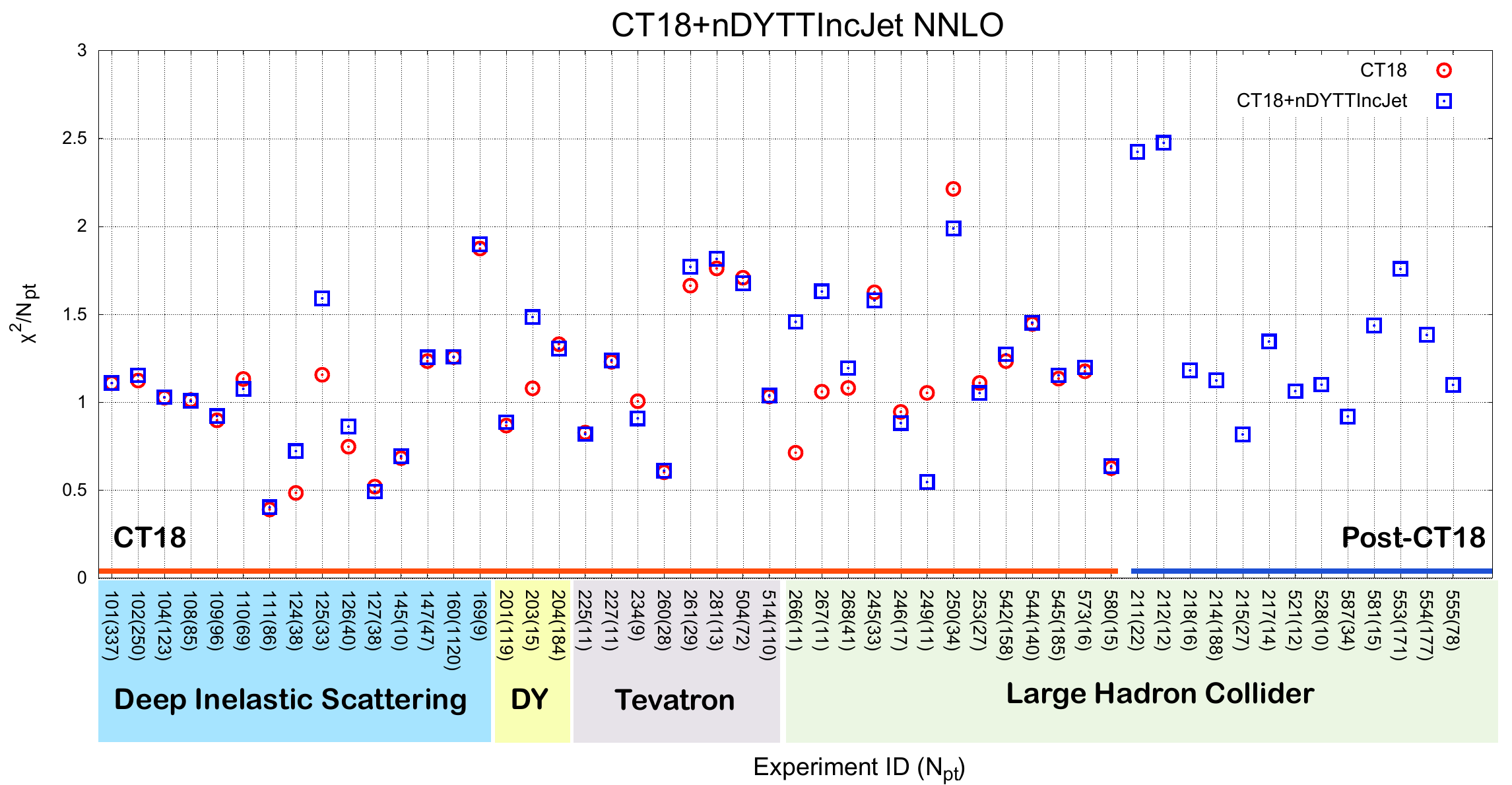}
	\caption{The $\chi^2/N_{\rm pt}$ values for each data set included in CT18+nDYTTIncJet NNLO fit, for both the resulting CT18+nDYTTIncJet PDFs and CT18 NNLO PDFs from \cite{Hou:2019efy}. The horizontal axis indicates the IDs and $N_{\rm pt}$ of the experiments, with the IDs in the CT18 baseline listed in \cite{Hou:2019efy} as well as  Table~\ref{Tab:LHCnewData}.}
	\label{Fig:Chi2Npt}
\end{figure}

For the combined analysis, the central question therefore concerns the selection of the data sets that are both constraining on the PDFs and consistent with one another. Based on the in-depth investigations in Refs.~\cite{Sitiwaldi:2023jjp,Ablat:2023tiy,Ablat:2024Jet}, we identified such an optimal selection, in which 13 distributions from the three categories of processes (for production of lepton pairs, top-quark pairs, and single-inclusive jets), with a total of 776 new data points, were added to the CT18 NNLO baseline fit, which contains a total of 3681 data points. This extension of the CT18 global data is named ``CT18+nDYTTIncJet''. The corresponding fit used the same underlying parametrization and analysis settings as the CT18 NNLO fit. Table~\ref{Tab:LHCnewData} lists  the newly added LHC data sets, together with the $\chi^2/N_{\rm pt}$ values obtained in the ``CT18+nDYTTIncJet'' NNLO fit and using the original CT18 NNLO PDFs (in parentheses). Figure~\ref{Fig:Chi2Npt} illustrates the $\chi^2/N_{\rm pt}$ values for all experiments in the CT18+nDYTTIncJet study by plotting them against the numerical ID and number of points of each experiment. 

\begin{wrapfigure}{r}{0.45\textwidth}
\vspace{-0.8cm}
 \includegraphics[width=0.45\textwidth]{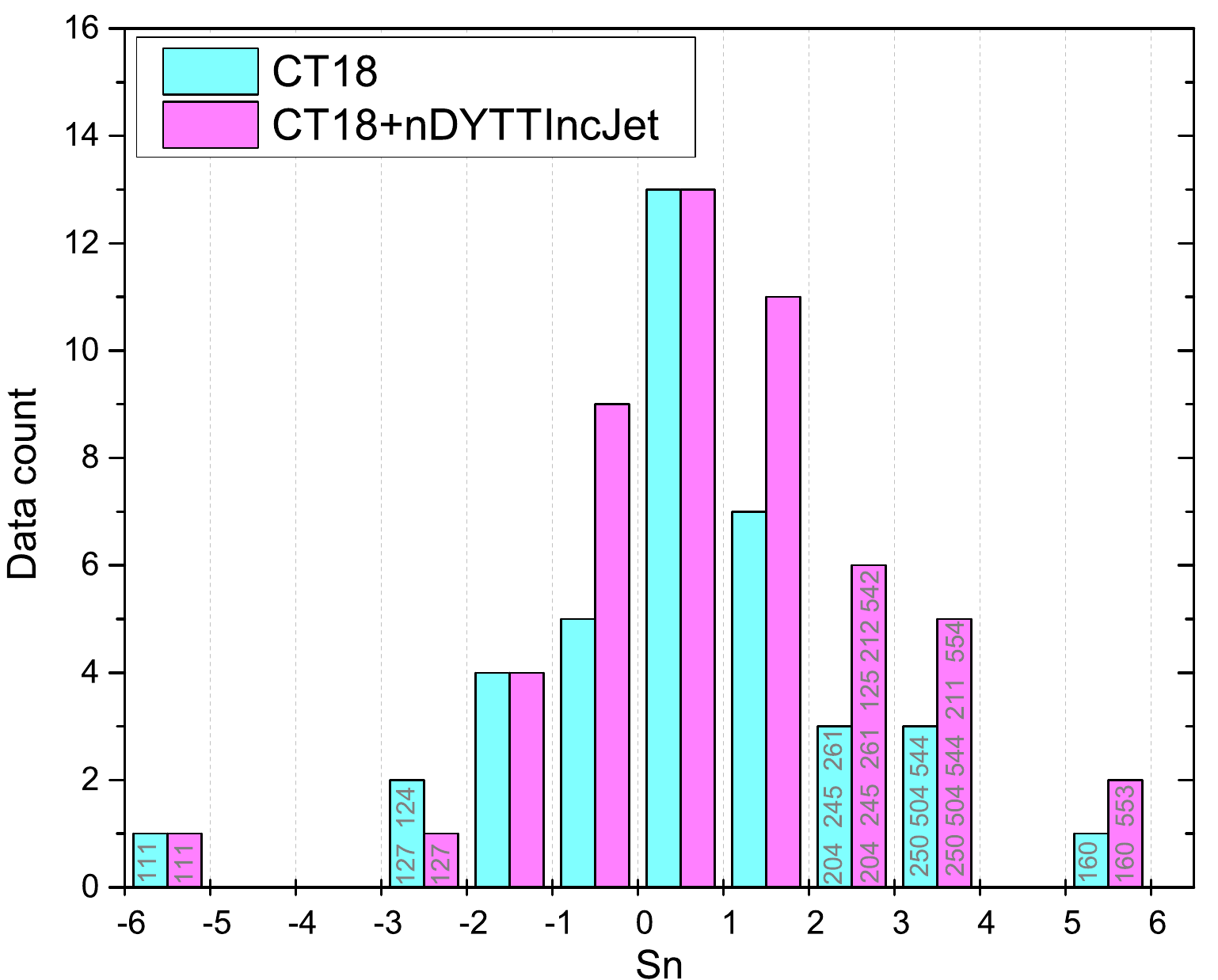}
	\caption{A histogram of the effective Gaussian variable, $S_n \equiv \sqrt{2\chi^2_n} - \sqrt {2 N_{\rm pt} -1}$,
		distributed over all CT18 and CT18+nDYTTIncJet data sets.}
	\label{Fig:Sn}
\end{wrapfigure}

The CT18+nDYTTIncJet fit is in good overall agreement with the old and new data sets, with the total $\chi^2/N_{\rm pt}=5402/4457\approx 1.21$ comparable to that in the CT18 NNLO baseline fit. We also observe some tensions among the data sets, which reduce their net constraining power. For example, in Fig.~\ref{Fig:Chi2Npt}, $\chi^2/N_{\rm pt}$ increases for the NuTeV (124 and 125), CCFR(126 and 127) and E866 DY (203) experiments, and the CMS 7 TeV muon charge asymmetry and electron charge asymmetry data sets (266  and 267). 
$\chi^2/N_{\rm pt}$ decreases for the CMS 8 TeV muon charge asymmetry (249) and LHCb 8 TeV  $W/Z$ (250) experiments. 

The overall level of tensions remains about the same as in the CT18 NNLO baseline fit, as revealed by the histogram of the effective Gaussian variable~\cite{Lai:2010vv}, defined as $S_n \equiv \sqrt{2\chi^2_n} - \sqrt {2 N_{\rm pt} -1}$, for individual data sets in the CT18 and CT18+nDYTTIncJet fits shown in Fig.~\ref{Fig:Sn}.
The $S_n$ values do not follow the standard normal distribution as would be expected in an ideal fit \cite[Sec. 4H]{Kovarik:2019xvh}. 
The $S_n$  values of the DIS HERA (160) and  ATLAS 8TeV IncJet (553) data sets are exceptionally large, while the neutrino-iron DIS CCFR F2 (111) data have a very good $S_n$  value.

\begin{figure}[tb]
	\includegraphics[width=0.49\textwidth]{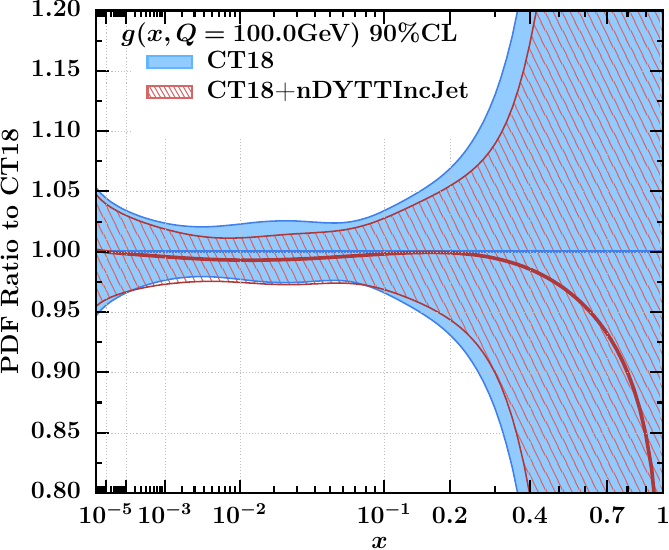}
   \includegraphics[width=0.49\textwidth]{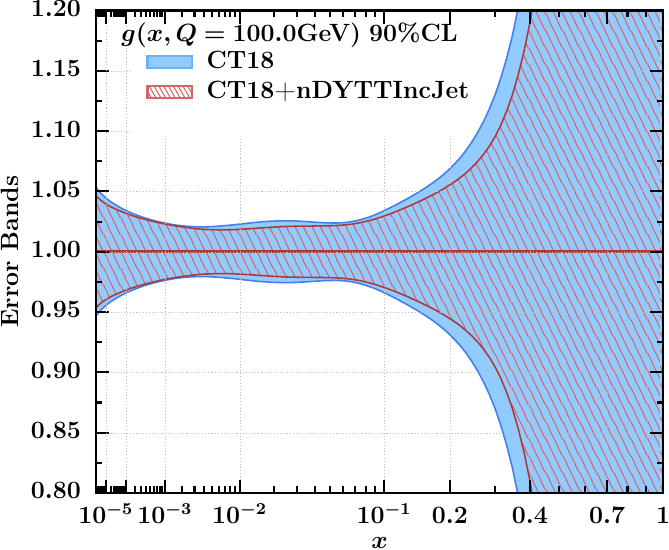}
	\caption{Comparisons of the gluon PDFs in the CT18 (blue) and CT18+nDYTTIncJet (red) NNLO PDF fits. Left: plotted as a ratio to the central CT18 NNLO PDF. Right: plotted as normalized error bands at 90\% CL.}\label{Fig:CT18+nDYTTIncJet}
\end{figure}

Figure~\ref{Fig:CT18+nDYTTIncJet} illustrates the combined impact of these new LHC Drell-Yan, $t\bar t$ and inclusive jet data sets on the gluon PDF at $Q=100$~GeV. In the left panel, the CT18+nDYTTIncJet fit prefers a softer gluon PDF at $x > 0.3$. Specifically, the downward pull on the large-$x$ gluon by the included Drell-Yan and $t\bar t$ data sets (with the latter given by the ``nTT2'' combination of $t\bar t$ observables introduced in \cite{Ablat:2023tiy}) overcomes the net upward pull from single-inclusive jet data sets. We also observe reduction of order 1\% in the gluon at moderate $x$ relevant for Higgs boson production at the LHC.  In the right panel, we see that the nominal CT18 uncertainty on the gluon is reduced across most of the $x$ range upon addition of the nDYTTIncJet combination. In this case, the uncertainties on the blue and red error bands are estimated according to the same prescription (tolerance) as in \cite{Hou:2019efy}. We note that the tolerance depends on methodology  and may result in different uncertainty estimates even when fitting the same data sets \cite{PDF4LHCWorkingGroup:2022cjn}. Advancing toward the near-future CTEQ-TEA analysis, we expect the PDF tolerance to evolve, on the one hand, to account for the residual tensions that remain significant in the CT18+nDYTTIncJet fit, cf. Fig.~\ref{Fig:Sn}, and, on the other hand, to quantify the dependence on PDF parametrization forms along the possibilities explored e.g. in \cite{Kotz:2023pbu,Kriesten:2023uoi, Kriesten:2024are}. Further details on the new CTEQ-TEA precision fit and its phenomenological implications will be presented in our future publication. 

\small

\section*{Acknowledgments}

The work of AA, SD and IS was supported by the National Natural Science
Foundation of China under Grants No.11965020 and No. 11847160. The
work of T.-J. Hou was supported by Natural Science Foundation of Hunan
province of China under Grant No. 2023JJ30496.  AC was supported by the
UNAM Grant No. DGAPA-PAPIIT IN111222 and CONACyT Ciencia de Frontera
2019 No. 51244 (FORDECYT-PRONACES).
MG was supported by the National Science Foundation under Grant
No.~PHY-2112025 and No.~PHY-2412071.
The work of T.~J.~Hobbs at Argonne National Laboratory was supported
by the U.S.~Department of Energy, Office of Science, under Contract
No.~DE-AC02-06CH11357.
PMN was partially supported by the U.S. Department of Energy under
Grant No.~DE-SC0010129.
The work of CPY and KX was supported by the U.S. National Science Foundation under
Grant No.~PHY-2310291. KX was also supported by the U.S. National
Science Foundation under Grant No.~PHY-PHY-2310497.
The work of PN and KX was performed in part at the Aspen Center for Physics,
which is supported by National Science Foundation grant PHY-2210452.
This work used resources of high-performance computing clusters from
SMU M2/M3, MSU HPCC, KSU HPC, as well as Pitt CRC.


\end{document}